\documentstyle[anta,twoside]{article}

\input{psfig.sty}

\def\Journal#1#2#3#4{(#1) {#2} {\bf #3}, #4}

\def\AAp{\em Astron. Astrophys.}

\def\ApJ{\em Astrophys.~J.}

\def\Arep{\em Astron. Rep.}
\def\MNRAS{\em Mon. Not. R.~Astron. Soc.}

\newcommand{\HII}{{\rm H\,\scriptstyle II}}
\begin{document}

\markboth{E.N. Vinyajkin}{Faraday Rotation and Depolarization of
Galactic Radio Emission}

\thispagestyle{plain}
\setcounter{page}{93}

\title{
Faraday Rotation and Depolarization of Galactic Radio Emission in the
Magnetized Interstellar Medium
}

\author{E.\,N. Vinyajkin}

\address{
Radiophysical Research Institute (NIRFI),\\
25 B.\,Pecherskaya st.,
Nizhny Novgorod,  603950, Russia
}

\maketitle

\abstract{ A joint action of depth and bandwidth depolarization in the
interstellar medium is considered using a model of $N$ homogeneous
synchrotron layers with Faraday rotation. The bandwidth depolarization
can be used in multifrequency polarimetric observations of Galactic
diffuse synchrotron radio emission to investigate the interstellar
ionized medium and magnetic field in the direction to the Faraday-thick
objects of known distances.}

\section{Introduction}

Faraday rotation and depolarization have considerable impact on
the angular pattern and frequency dependence of the position angle and
brightness temperature of the linearly polarized component of the
diffuse Galactic synchrotron radio emission. This effect increases
with the distance from which we receive the linearly polarized radio
emission. The observing frequency, bandwidth and beamwidth play
important roles. Faraday depolarization may be caused by:
1)~differential Faraday rotation along the line of sight when
synchrotron emission and Faraday rotation are mixed (depth or
front-back depolarization), 2)~differential Faraday rotation in the
receiver  bandwidth (bandwidth depolarization), and 3)~difference of
Faraday rotation (and also intrinsic position angles) within the
beamwidth (beamwidth depolarization). Depth and bandwidth
depolarizations act at sufficiently low frequencies even in the case of
a homogeneous radiation region and infinitely narrow antenna beam. Here
we consider a joint action of depth and bandwidth depolarization in the
interstellar medium using simple models of the regions with the
synchrotron radio emission and Faraday rotation. We assume that
1)~the receiver bandwidth $\Delta\nu\ll\nu_0$  ($\nu_0$ is the central
frequency), 2)~the receiver frequency response is rectangular
\begin{equation}\label{v1}
\begin{array}{l}
\displaystyle
{\cal F}(s)=1,\hbox{ \ if \ } |s|\le{\Delta\nu\over2\nu_0}\,,\\[3mm]
\displaystyle
{\cal F}(s)=0,\hbox{ \ if \ } |s|>{\Delta\nu\over2\nu_0}\,,
\end{array}
\end{equation}
where $s= (\nu-\nu_0)/\nu_0$, 3)~the beam is narrow enough to neglect
the difference between position angles of waves coming from different
directions.

\section{Depth and bandwidth depolarization}

\subsection{A homogeneous region behind the Faraday screen}

Let us consider a model consisting of a radio emission region of
extension $L$ along the line of sight with a homogeneous magnetic field
and homogeneous space distributions of relativistic and thermal electrons
and some other object located in front of it with substantial Faraday
rotation and nonpolarized or negligibly small self-emission.
Such an object can be an $\HII$ region, a magnetic bubble (Vall\'ee,
1984), a planetary nebula, an external part of a molecular cloud
(Uyan{\i}ker \& Landecker, 2002; Wolleben \& Reich, this volume), a
depolarized supernova remnant (SNR), the solar corona (Soboleva \&
Timofeeva, 1983; Mancuso \& Spangler, 2000), or the Earth ionosphere.
The near object is the Faraday screen for the region located behind it.
Stokes parameters $Q$ and $U$ in this model account for depolarization
in the rectangular bandwidth~(1) (Vinyajkin \& Krajnov, 1989)
\begin{equation}\label{v2}
\begin{array}{rcl}
Q&=&\displaystyle
{P_0I\over2\phi\lambda^2}\Big(
F\left[2(\phi+\phi_s)\,\lambda^2\delta\right]
\sin\!\left\{2\left[\chi_0+(\phi+\phi_s)\,\lambda^2\right]\!\right\}-\nonumber\\
&&-\,\displaystyle
F(2\phi_s\lambda^2\delta)\sin\!\left[
2(\chi_0+\phi_s\lambda^2)\right]\!\Big),\nonumber\\[2mm]
U&=&\displaystyle
{P_0I\over2\phi\lambda^2}\Big(
-F\left[2(\phi+\phi_s)\,\lambda^2\delta\right]
\cos\!\left\{2\left[\chi_0+(\phi+\phi_s)\,\lambda^2
\right]\!\right\}+\nonumber\\
&&+\,\displaystyle
F(2\phi_s\lambda^2\delta)\cos\!\left[
2(\chi_0+\phi_s\lambda^2)\right]\!\Big),
\end{array}
\end{equation}
where $P_0$  is the intrinsic polarization degree, $\chi_0$  is the
intrinsic position angle,  $I$ is the intensity, $\phi=0{.}81({\rm
rad}\cdot{\rm m}^{-2}{\rm cm}^3\mu{\rm G}^{-1} {\rm pc}^{-1})N_eB_\|L$
is the Faraday depth of the radiation region, $N_e$ is the electron
density, $B_\|={\bf Bk}/k$  is the component of the magnetic field $\bf
B$  along the line of sight (${\bf k}$  is the wave vector, $k=2\pi
/\lambda$ is the wave number), $\phi_s=0{.}81({\rm rad}\cdot{\rm
m}^{-2}{\rm cm}^3\mu{\rm G}^{-1} {\rm pc}^{-1}){(N_e)}_sB_{\| s}L_s$ is
the Faraday screen depth, $\delta=\Delta\nu /\nu$,  and
\begin{equation}\label{v3}
F(2x\lambda^2\delta)={\sin(2x\lambda^2\delta)\over2x\lambda^2\delta}\,.
\end{equation}

\subsection{ $N$ different homogeneous layers}

Now let us consider a more general model consisting of $N$
different homogeneous layers (Sokoloff et al., 1998), each of them
characterized by three parameters: the intensity $I_i$,  Faraday depth
$\phi_i$,  and intrinsic position angle $\chi_{0i}$, where
$i=1,2,\ldots,N$  and the farthest region has $i=1$.  If one of the
layers is not a source of linearly polarized radio emission and only
rotates the polarization plane, then  $I_i=0$. The intrinsic
polarization degree of any emitting layer is the same and equals to
$P_0$.

Expressions for Stokes parameters of the $N$-layer model are easily
obtained from~(2) and (3), if we take into account that all
regions from $i+1$ up to $N$ play the role of the Faraday screen for the
$i$-region and rotate the polarization plane by the angle
$\displaystyle\left(\sum\limits_{j=i+1}^N\phi_j\right)\lambda^2$.
Because the Stokes parameters are additive for noncoherent radio
emission, the values of $Q_N$, $U_N$  for the $N$-layer region can
be obtained by summing up over all $N$ components~(Vinyajkin et al., 2002):
\begin{eqnarray}\label{v4}
Q_N=P_0\sum_{i=1}^N\,{I_i\over2\phi_i\lambda^2}&&\hspace{-7mm}
\left(
F\left[2\left(\phi_i+\sum_{j=i+1}^N\phi_j\right)
\lambda^2\delta\right]\!\sin\!\left\{2\!\left[
\chi_{0i}+\left(\phi_i+\sum_{j=i+1}^N\phi_j\right)
\lambda^2\right]\!\right\}-\right.\nonumber\\
&&-\left.
F\left[2\left(\sum_{j=i+1}^N\phi_j\right)
\lambda^2\delta\right]\!\sin\!\left\{2\!\left[
\chi_{0i}+\left(\sum_{j=i+1}^N\phi_j\right)
\lambda^2\right]\right\}\right)\!,\nonumber\\
\\
U_N=P_0\sum_{i=1}^N\,{I_i\over2\phi_i\lambda^2}&&\hspace{-7mm}
\left(
{-}F\!\left[2\left(\phi_i+\!\sum_{j=i+1}^N\phi_j\right)
\lambda^2\delta\right]\!\cos\!\left\{2\!\left[
\chi_{0i}+\!\left(\phi_i+\sum_{j=i+1}^N\phi_j\right)\!
\lambda^2\right]\!\right\}\!+\right.\nonumber\\
&&+\left.
F\left[2\left(\sum_{j=i+1}^N\phi_j\right)
\lambda^2\delta\right]\!\cos\!\left\{2\!\left[
\chi_{0i}+\left(\sum_{j=i+1}^N\phi_j\right)
\lambda^2\right]\right\}\right)\!.\nonumber
\end{eqnarray}
If $\delta\to0$, $F\to1$,  and Eqs. (4) correspond to
eq. (9) from Sokoloff et al.\ (1998).

\subsubsection{$N=1$}

In the case of a single homogeneous layer ($N=1$) we get from~(4):
\begin{eqnarray}\label{v5}
&&
Q_1=P_0\,{I\over2\phi\lambda^2}\,\left(
{\sin(2\phi\lambda^2\delta)\over2\phi\lambda^2\delta}
\sin\!\left[2\left(\chi_0+\phi\lambda^2\right)\!\right]-
\sin2\chi_0\right)\!,\nonumber\\
\\
&&
U_1=P_0\,{I\over2\phi\lambda^2}\,\left(-
{\sin(2\phi\lambda^2\delta)\over2\phi\lambda^2\delta}
\cos\!\left[2\left(\chi_0+\phi\lambda^2\right)\!\right]+
\cos 2\chi_0\right)\!.\nonumber
\end{eqnarray}
The depolarization factor $DP=\sqrt{Q^2+U^2}/IP_0=P/P_0$, where $P$ is
the observed degree of polarization,  is equal
to~(Vinyajkin \& Krajnov, 1989; Vinyajkin \& Razin, 2002)
\begin{equation}\label{v6}
DP_1={1\over2\,|\phi|\,\lambda^2}\left(
{\left[{\sin(2\phi\lambda^2\delta)\over2\phi\lambda^2\delta}
\right]}^2-2\,{\sin(2\phi\lambda^2\delta)\over
2\phi\lambda^2\delta}\cos(2\phi\lambda^2)+1\right)^{1/2}\!.
\end{equation}
The dotted lines in Figs.~1 and 2 represent the dependencies of
the depolarization factor~(6) on $\phi\lambda^2/2$  in the intervals
$0\div10$ and $70\div25\pi$, respectively, for the typical value
$\delta=0{.}01$. The solid lines in these figures show the
dependencies of the observed position angle  $\chi_{1\,{\rm obs}}$ on
$\phi\lambda^2/2$\footnote{
\,Here the observed values of the position angle are those measured in
the interval $0\div\pi$ and counted counter-clockwise from the vertical.
In Figs.~1 and 2 the observed position angles are identical with the
true ones. In the general case the observed value of the position angle
may differ from the true one by $\pm n\pi$ $(n=0, 1, 2,\dots)$.
}
\begin{equation}\label{v7}
\chi_{1{\rm obs}}={1\over2}\,{\rm angle}\,(Q_1,U_1),
\end{equation}
where ${\rm angle}\,(x, y)$ gives the angle in radians between the axis
$x$ (vertical axis $Q$) and the vector with coordinates $x$, $y$
(polarization vector on the plane $Q$, $U$). It is seen from Eqs.
(5) and (6) and from Figs.~1 and 2 that the oscillation amplitude of the
position angle near the value $(\chi_0+\pi/2)/2=\pi/4$ (assuming
$\chi_0=0$) decreases with increasing $\phi\lambda^2/2$, and at
$\phi\lambda^2/2=\pi/4\delta=25\pi$  the position angle
$\chi_{1\,{\rm obs}}=\pi/4$, $DP_1=\delta/\pi=(1/\pi)\%$.

\begin{figure}[p]
\begin{minipage}[t]{7.5cm}
\psfig{figure=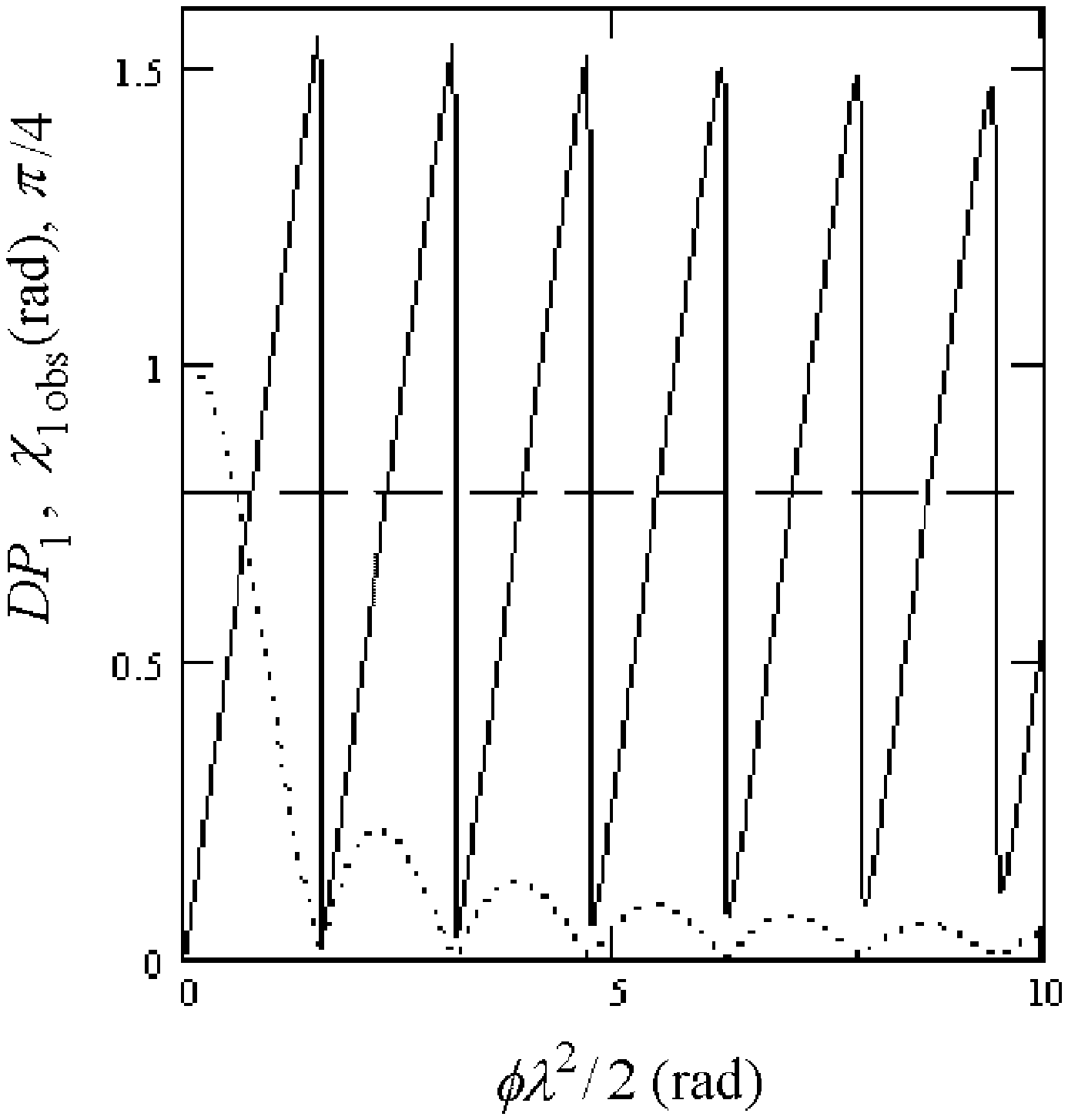,width=7.5truecm}
\caption{The depolarization factor $DP_1$ (dotted line),
the observed position angle $\chi_{1\,{\rm obs}}$ (solid line),
and the number $\pi/4$ (dashed line), see text.}
\end{minipage}\hfill
\begin{minipage}[t]{7.5cm}
\psfig{figure=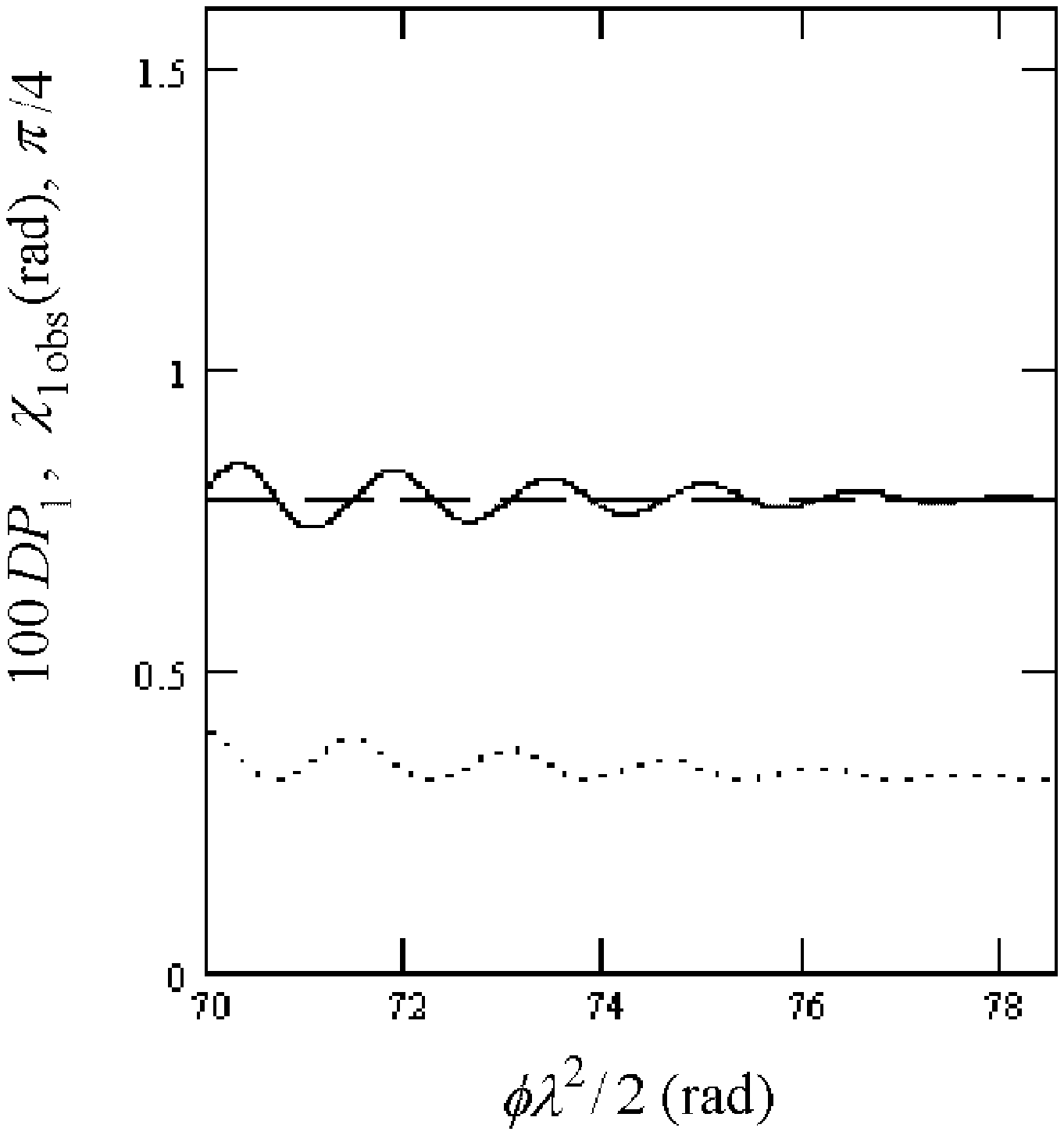,width=7.5truecm}
\caption{Same as Fig.\,1, but for the $\phi\lambda^2/2$
interval 70 to 25$\pi$ instead of 0 to 10, and 100 $DP_1$
instead of $DP_1$.}
\vspace{1.5cm}
\end{minipage}
\begin{minipage}[t]{7.5cm}
\psfig{figure=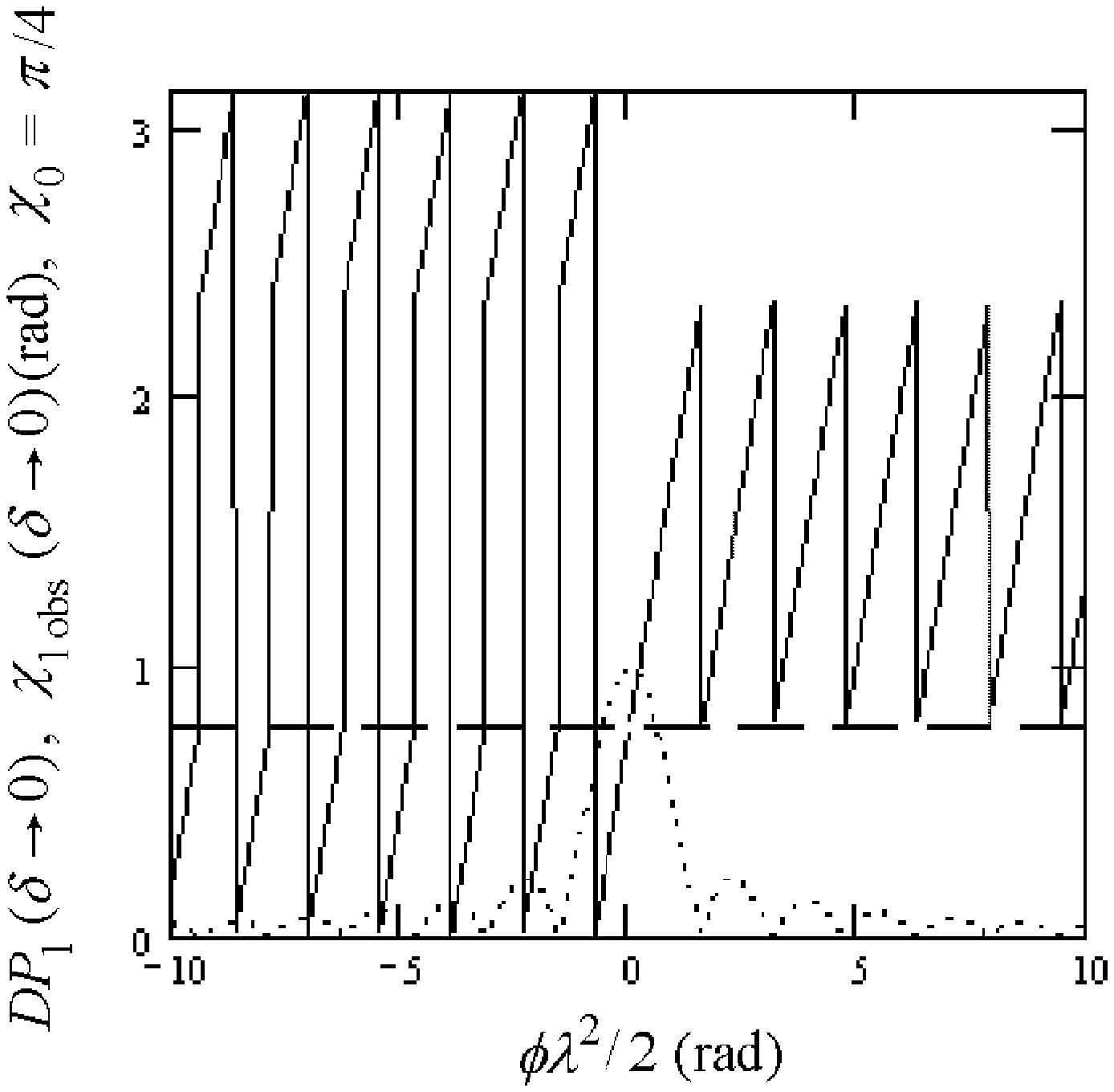,width=7.5truecm}
\caption{The depolarization factor $DP_1$ $(\delta\to0)$ (dotted line),
the observed position angle $\chi_{1\,{\rm obs}}$ $(\delta\to0)$
(solid line), and $\chi_0=\pi/4$ (dashed line).}
\end{minipage}\hfill
\begin{minipage}[t]{7.5cm}
\psfig{figure=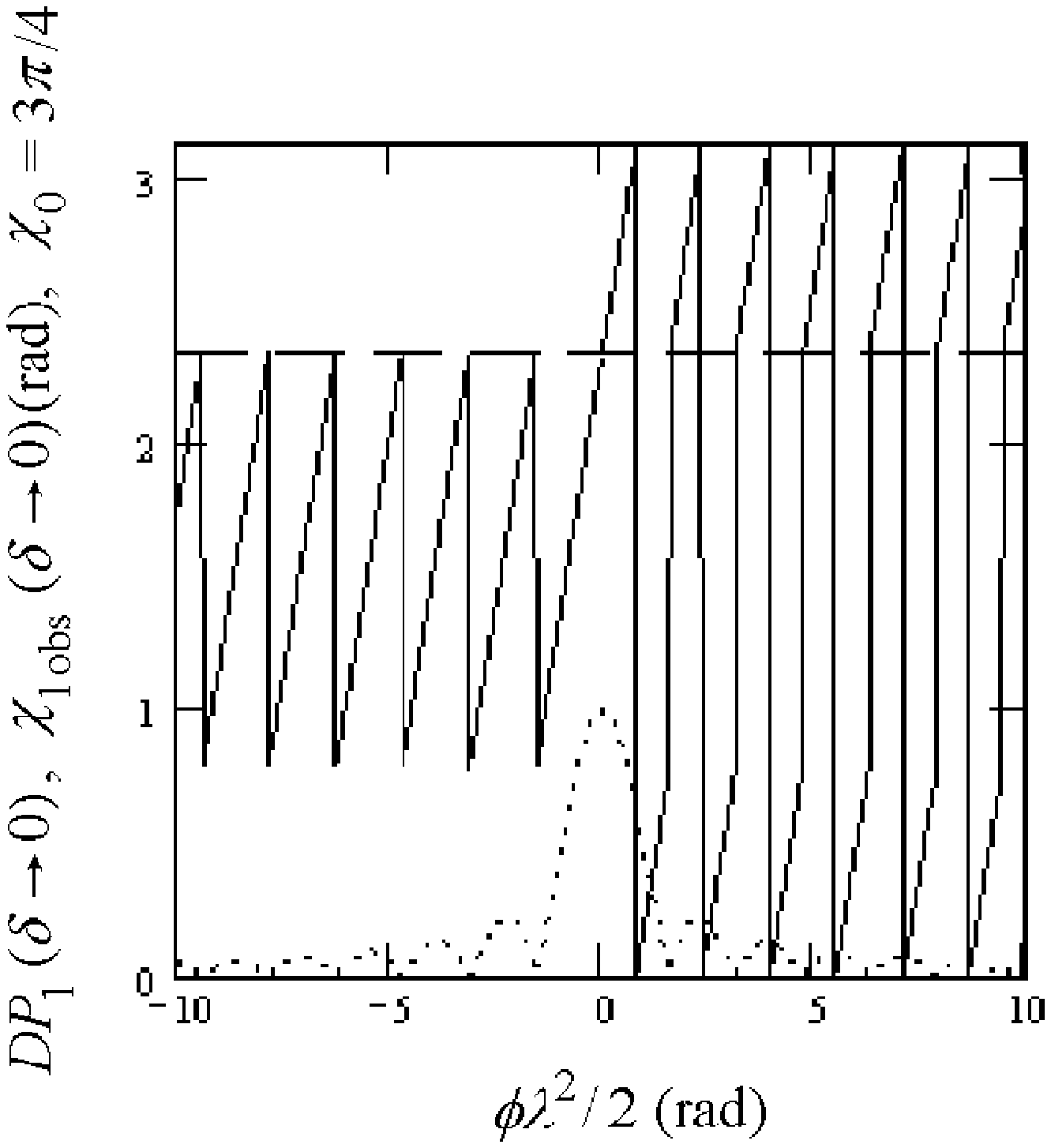,width=7.5truecm}
\caption{Same as Fig.\,3, but for $\chi_0=3\pi/4$. }
\end{minipage}
\end{figure}

 In the limit of infinitely narrow band $\delta\ll1/2\,|\phi|\,\lambda^2$
(we have assumed $\delta\ll1$) Eqs. (5) transform to
\begin{eqnarray}\label{v8}
&&
Q_1(\delta\to0)=P_0I\,{\sin\phi\lambda^2\over\phi\lambda^2}
\cos\left[2\left(\chi_0+{\phi\over2}\,\lambda^2\right)\!\right]\!,
\nonumber\\
\\
&&
U_1(\delta\to0)=P_0I\,{\sin\phi\lambda^2\over\phi\lambda^2}
\sin\left[2\left(\chi_0+{\phi\over2}\,\lambda^2\right)\!\right]\!,
\nonumber
\end{eqnarray}
and (6) transforms into the known formula for the depolarization factor
of a homogeneous synchrotron layer with rotation in the limit of the
infinitely narrow bandwidth and beam~(Razin, 1956)
\begin{equation}\label{v9}
DP_1(\delta\to0)=\left|{\sin\phi\lambda^2\over\phi\lambda^2}\right|.
\end{equation}
The position angle corresponding to Stokes parameters~(8) equals
to~(Razin, 1956; Burn, 1966; Vinyajkin, 1995)
\begin{equation}\label{v10}
\chi_1(\delta\to0)=\chi_0+{\phi\over2}\,\lambda^2-
{\pi\over2}\,E(\phi\lambda^2/\pi),
\end{equation}
where $E(x)=-E(-x)$ is the integral part of argument $x$. The position
angle values of~(10) may come out of the interval $0\div180^\circ$,
for example, if $\chi_0>\pi/2$  and  $\phi>0$. To calculate the observed
values $\chi_{1\,{\rm obs}}$  one has to use Eq.~(7) with the Stokes
parameters from (8). Figures 3 and 4 give plots of $\chi_{1\,{\rm obs}}$
$(\delta\to0)$ as dependent on  $(\phi/2)\,\lambda^2$ (solid lines)
for the values of $\chi_0$, respectively, $\pi/4$  and
$3\pi/4$ (dashed lines). The depolarization factor~(9) is shown by
dotted lines. The rotation measure $RM=\phi/2$ and peculiarities of
its determination in this model have been considered in detail by
Vinyajkin (1995). The model of a homogeneous layer was used by
Vinyajkin~(1995) to give the interpretation of a deep minimum of the
polarization brightness temperature and a $\pi/2$-jump of the position
angle observed in the North Polar Spur in the direction with coordinates
$\alpha_{1950}=16^{\rm h}\,48^{\rm m}$, $\delta_{1950}=14^\circ$
at 960~MHz~(Vinyajkin, 1995).

\subsubsection{$N=3$}

Let us consider the model of the region consisting of two homogeneous
synchrotron polarized layers and the Faraday screen between them.
In this case the contribution of the far layer $(i=1)$ in the observed
polarized radio emission becomes 0 at some relatively low frequency
because of its bandwidth depolarization. If the distance to the
Faraday screen is known, we can estimate the extension of the near
synchrotron layer. As an example, let us consider the following model
parameters: $I_1/I = I_3/I = 0.5$, $I_2=0$, $\chi_{01}=\chi_{03}=0$,
$\phi_1=\phi_3=0$, $\phi_2=100$~rad/m$^2$. The contribution of the far
layer becomes 0 and, hence, the depolarization factor becomes 0.5 (see
Figs. 5 and 6) at the minimum wavelength
\begin{equation}\label{v11}
\lambda_{\min}=\sqrt{\pi\over2}\,{1\over\sqrt{\phi_2\delta}}\,,
\end{equation}
which corresponds to the first zero of the function
$|\sin(\Delta\chi)/\Delta\chi|$, where
$\Delta\chi=2\phi_2\lambda^2\delta$ is the differential Faraday
rotation in the bandwidth. Substituting $\phi_2=100$~rad/m$^2$,
$\delta=0{.}01$ in~(11) we get $\lambda_{\min}\approx1{.}25$~m
($\nu_{\max}\approx240$~MHz). Equation~(11) can be used to
estimate the cut-off wavelength of the far layer if $\phi_1\ll\phi_2$.
Let us consider some objects. The $RM$ of SNR CTB 104A changes from
$\sim-80$~rad/m$^2$ in the southeast to $\sim+170$~rad/m$^2$ in the
northwest~(Uyan{\i}ker et al., 2002). Assuming $\phi_2\sim340$~rad/m$^2$,
$\delta=0{.}01$  we get from (11) $\lambda_{\min}\sim0{.}7$~m
($\nu_{\max}\sim430$~MHz). At this wavelength this part of the SNR is
nearly completely depolarized because of the depth depolarization
($P<0{.}4\%$). The $RM$ of the $\HII$ region S205 is
250~rad/m$^2$~(Mitra et al. 2003; Wielebinski \& Mitra, this volume).
In this case $\phi_2=250$~rad/m$^2$ and, if $\delta=0{.}01$,  we get
$\lambda_{\min}\approx0{.}8$~m ($\nu_{\max}\approx375$~MHz).
Gray et al. (1999) detected a strong beam and bandwidth
depolarization across the face of W3 and W4 and immediately near them
at 1420~MHz (30~MHz bandwidth).

\begin{figure}[htb]
\begin{minipage}[t]{7.5cm}
\psfig{figure=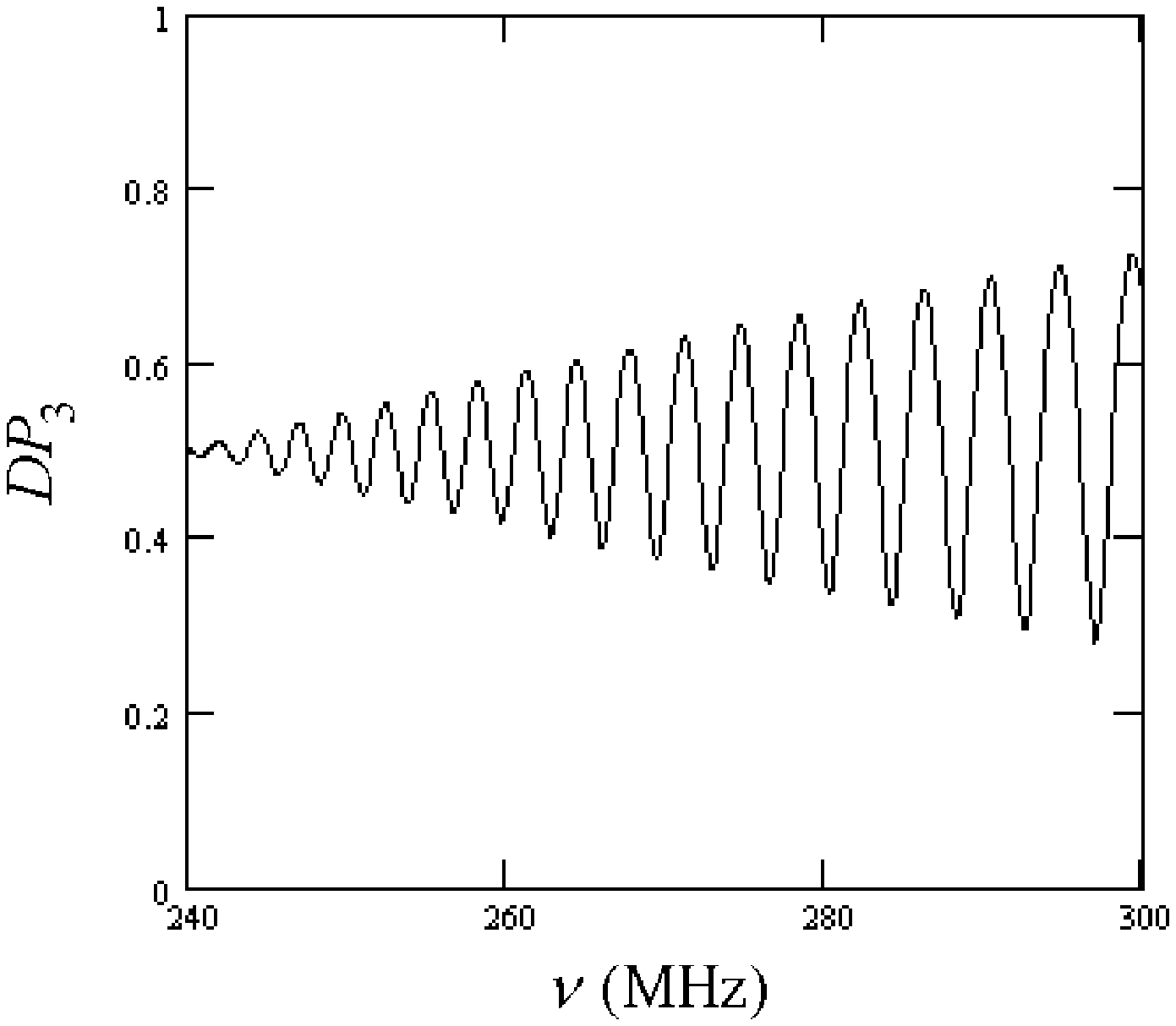,width=7.5truecm}
\caption{The depolarization factor $DP_3$ of the three-layer model
(see text) versus frequency  in the interval $240\div300$~MHz.  }
\end{minipage}\hfill
\begin{minipage}[t]{7.5cm}
\psfig{figure=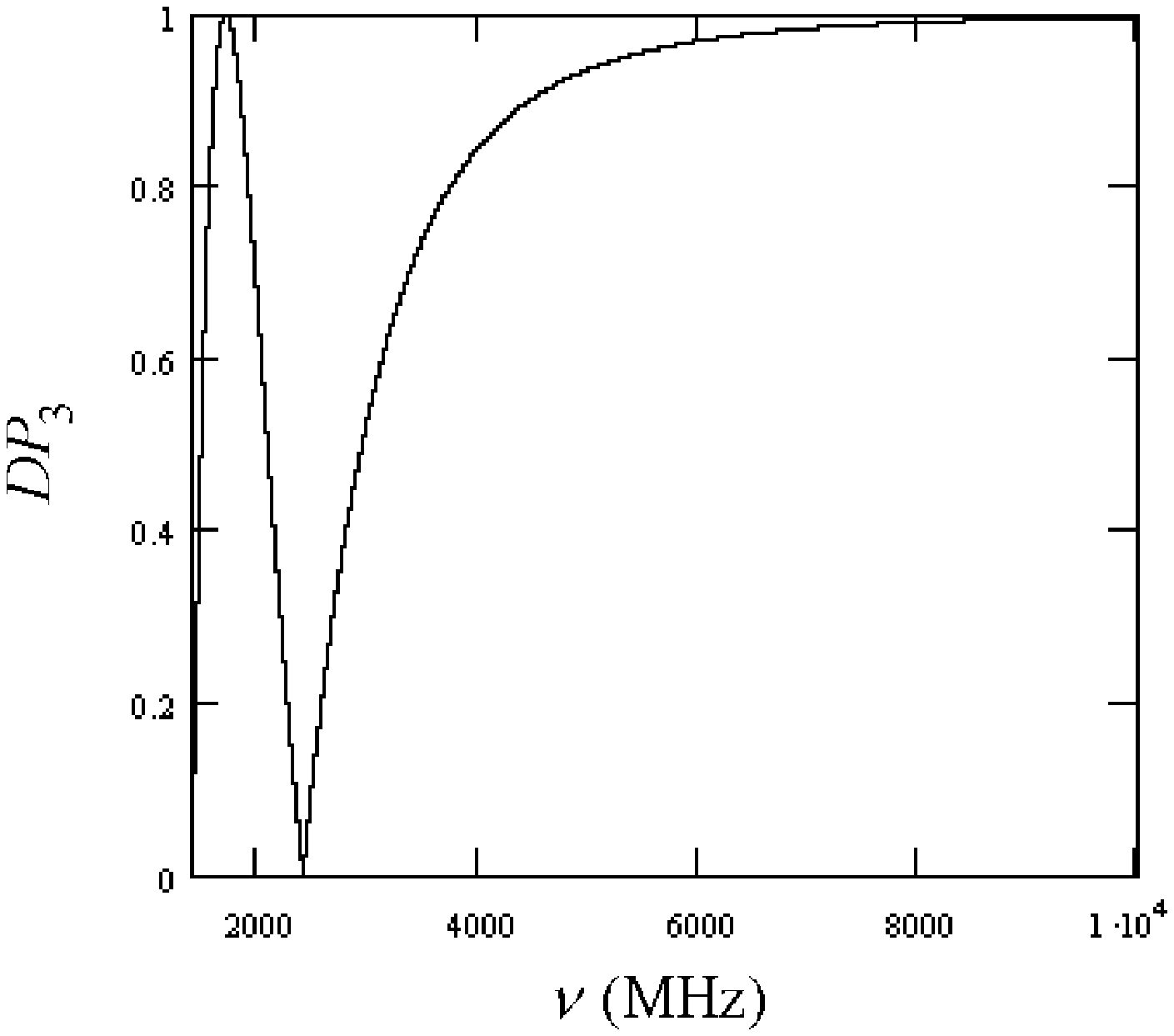,width=7.5truecm}
\caption{Same as Fig.\,5, but for the frequency  interval
$1400\div10000$~MHz.}
\end{minipage}
\end{figure}

Carrying out high angular resolution broadband polarimetric
multifrequency observations in the directions to the Faraday screens
with known distances it is possible to investigate the interstellar
ionized gas and magnetic field along the line of sight. However,
interference is a serious problem in carrying out such observations.

\section{Conclusion}

The bandwidth depolarization can be a useful tool in multifrequency
polarimetric observations of Galactic diffuse synchrotron radio
emission to investigate the interstellar ionized gas and magnetic field
in the directions to Faraday-thick objects of known distances.

\section*{Acknowledgment}

This work has been supported by the International Science and
Technology Center under the ISTC project No.~729.

\section*{References}\noindent

\references

Burn, B.\,J. \Journal{1966}{\MNRAS}{133}{67}.

Gray, A.\,D., Landecker, T.\,L., Dewdney, P.\,E.,
  Taylor,~A.\,R., Willis,~A.\,G.,
  Normandeau, M. \Journal{1999}{\ApJ}{514}{221}.

Mancuso, S., Spangler, S.\,R. \Journal{2000}{\ApJ}{539}{480}.

Mitra, D., Wielebinski, R., Kramer, M., Jessner, A.
  \Journal{2003}{\AAp}{398}{993}.

Razin, V.\,A. (1956) {\em Rad.\ i Elek.} {\bf 1}, 846.

Soboleva, N.\,S., Timofeeva, G.\,M. (1983) {\em Soviet Astron.\ Lett.}
  {\bf 9}, 216.

Sokoloff, D.\,D., Bykov,~A.\,A., Shukurov~A.,
  Berkhuijsen,~E.\,M., Beck,~R., Po\-ezd,~A.\,D.
  \Journal{1998}{\MNRAS}{299}{189}.

Uyan{\i}ker, B., Landecker, T.\,L. (2002) in {\em Astrophysical
  Polarized Backgrounds}, eds.\ S.\,Cecchini, S.\,Cortiglioni,
  R.\,Sault, \& C.\,Sbarra, AIP Conf.\ Proc.\ 609, (Melville: AIP), p.~15.

Uyan{\i}ker, B., Kothes, R., Brunt, C.\,M. \Journal{2002}{\ApJ}{565}{1022}.

Vall\'ee, J.\,P. \Journal{1984}{\AAp}{136}{373}.

Vinyajkin, E.\,N., Krajnov, I.\,L. (1989) Preprint of the
  Radiophysical Research Inst. (NIRFI), Gorky, No.\,288.

Vinyajkin, E.\,N. \Journal{1995}{\Arep}{39}{599}.

Vinyajkin, E.\,N., Razin, V.\,A. (2002) in {\em Astrophysical Polarized
   Backgrounds}, eds.\ S.\,Cecchini, S.\,Cortiglioni, R.\,Sault, \&
   C.\,Sbarra, AIP Conf.\ Proc.\ 609, (Melville: AIP), p.~26.

Vinyajkin, E.\,N., Paseka, A.\,M., Teplykh, A.\,I. (2002)
  {\em Radiophysics and Quantum Electronics} {\bf 45}, 102.

\end{document}